\documentclass[showpacs,preprint,preprintnumbers,showpacs,showkeys,superscriptaddress,amsmath,amssymb,nofootinbib]{revtex4}

\usepackage{amsmath,latexsym}
\def\[{\left\lbrack}
\def\]{\right\rbrack}

\def\({\left(}
\def\){\right)}
\def\ni{\noindent}
\newcommand{\be}{\begin{equation}}
\newcommand{\ee}{\end{equation}}
\newcommand{\ea}{\end{eqnarray}}
\newcommand{\ba}{\begin{eqnarray}}

\begin{document}

\title{Compressible Fluids with Maxwell-type equations, \\ the minimal coupling with electromagnetic field and \\  the Stefan-Boltzmann law}

\author{Albert C. R. Mendes}\email{albert@fisica.ufjf.br}
\affiliation{Departamento de F\'{i}sica, Universidade Federal de Juiz de Fora, 36036-330, Juiz de Fora - MG, Brazil}
\author{Flavio I. Takakura}\email{takakura@fisica.ufjf.br}
\affiliation{Departamento de F\'{i}sica, Universidade Federal de Juiz de Fora, 36036-330, Juiz de Fora - MG, Brazil}
\author{Everton M. C. Abreu}\email{evertonabreu@ufrrj.br}
\affiliation{Grupo de F\' isica Te\'orica e Matem\'atica F\' isica, Departamento de F\'{i}sica, Universidade Federal Rural do Rio de Janeiro, 23890-971, Serop\'edica - RJ, Brazil}
\affiliation{Departamento de F\'{i}sica, Universidade Federal de Juiz de Fora, 36036-330, Juiz de Fora - MG, Brazil}
\author{Jorge Ananias Neto}\email{jorge@fisica.ufjf.br}

\date{\today}



\begin{abstract}
\ni In this work we have obtained a higher-derivative Lagrangian for a charged fluid coupled with the electromagnetic fluid and the Dirac's constraints analysis was discussed.   A set of first-class constraints fixed by noncovariant gauge condition was obtained.   The path integral formalism was used to obtain the partition function for the corresponding higher-derivative Hamiltonian and the Faddeev-Popov ansatz was used to construct an effective Lagrangian.   Through the partition function, a Stefan-Boltzmann type law was obtained.
\end{abstract}

\maketitle

\section{Introduction}

In recent papers the authors have discussed that, as an alternative way for the description of fluid dynamics, concerning both the compressible fluids \cite{Kambe} and the equations of  plasma \cite{Thompson}, the better path would be through the recasting of the equations of motion to obtain a set of Maxwell-type equations for the fluid. This transformation in the structure of the equations of motion results in the generalization of the concept of charge and current connected to the dynamics of the fluid \cite{Marmanis,jnpp}.  The identification of what will be considered as a source term in the resulting theory depends on the choice of the objects which will form the main part of its new structure of fluid dynamics. In Lighthill's work concerning the sound radiated by a fluid flow \cite{lighthill}, the applied stress tensor was considered as the source of the radiation field. R. J. Thompson \cite{Thompson} recently introduced an extension of this new structure of the plasma equations of motion, for each kind of fluid, from the equations of motion that  describe such system. 

The reason is to understand thermodynamical arguments in order to obtain how the energy density $\rho$ depends on the temperature $T$ for a fluid's equation of state given by $p=\omega \rho$.  Besides, the Stefan-Boltzmann law has been widely discussed in the scenario of black holes thermodynamics 
\cite{hawking}, from where we know that the energy density is inversely proportional to the temperature.  More recently, the observed acceleration of the Universe demands the existence of a new component, the termed dark energy, which rules out all other forms of energy and has a negative  pressure.  The presence of such energy in the Universe deserves detailed analysis, such as the consequences related to the application of the generalized second law \cite{beckenstein} or the entropy bound \cite{beckenstein2}.   Some elements, such as the phantom field ($\omega < -1$), fields with a negative kinetic energy, negative temperature and the entropy being always positive, can change completely the evolution of black holes and their connection to the generalized second law, as was discussed in \cite{fh,gs}.

With these motivations in mind, our proposal here, besides to consider a charged fluid, which is defined by the Lagrangian density (\ref{0.1}) below, is to discuss its coupling with the electromagnetic field and its corresponding constraints through Dirac's constraint classification. Another important point will be the construction of the theory partition function  after determining its constraint structure \cite{Albert1}, which allows us, finally, to evaluate all the thermodynamical quantities. 

The work is organized in such a way that in section 2 we have a description of the general aspects of the theory's canonical structure. In section 3 we present the transition amplitude using the path integral formulation. In section 4 we will analyze the theory in thermodynamical equilibrium using the imaginary-time formalism and we will derive the Stefan-Boltzmann law and finally in the last section we present the conclusions.

\section{Canonical Structure}

Recently \cite{albert1}, some of us have shown that a Lagrangian formulation for a compressible fluid can be obtained, analogously to the one described by
Marmanis concerning an incompressible fluid \cite{Marmanis}, resulting in a Maxwell-type action for the fluid considering the viscosity, given by
\be\label{0.1}
{\cal L}_{fluid} = -{{1}\over{4}} T_{\mu\nu}T^{\mu\nu}\,\,,
\ee
where $T_{\mu\nu} =\partial_\mu U_{\nu} -\partial_\nu U_{\mu}$ is the strength tensor of the fluid.  The four-vector potential $U_\mu \equiv (U_0, \vec{U})$, where $U_0$ is the energy function and $\vec {U}$ is the average velocity field \cite{albert1}. The spacetime metric is $\eta_{\mu\nu}=(-+++)$.

The Lagrangian density in Eq. (\ref{0.1}) gives us the set of  Maxwell-type equations for the homogeneous case (no sources) \cite{albert1}, from which we can derive the main equations in fluid dynamics that are the equations for the vortex dynamics $\vec \omega$
\be\label{0.2}
{{\partial\vec\omega}\over{\partial t}} + \nabla \times(\vec\omega \times \vec U) =\nabla \times \vec z \,\,,
\ee
where  $\vec z = T\nabla s + \rho^{-1}\nabla \sigma$, $T$ is the temperature, $s$ is the entropy per unit mass and $\rho$ is the fluid density.  And $\sigma$ is the stress tensor \cite{Landau}. 

The right term in Eq. (\ref{0.2}) that can be rewritten as
\be\label{0.3}
\nabla\times\vec z = \nabla T \times \nabla s + \nabla\times(\rho^{-1}\nabla\sigma) = \vec{\Gamma}_B + \vec{\Gamma}_\nu \,\,,
\ee
therefore, we have that
\be\label{0.4}
{{\partial\vec\omega}\over{\partial t}} + \nabla \times(\vec\omega \times \vec U) = \vec{\Gamma}_B + \vec{\Gamma}_\nu \,\,,
\ee
which has, precisely, the standard form. In the above Eqs. \eqref{0.3}-\eqref{0.4}, $\vec{\Gamma}_B$ and $\vec{\Gamma}_\nu$, explicitly displayed on the right hand side, are the possible sources of the vorticity $\vec \omega$, where
\be\label{0.5}
\vec{\Gamma}_B =\nabla T \times \nabla s
\ee
that is equivalent to the  traditional Biermann battery \cite{biermann}, and the second term
\be\label{0.6}
\vec{\Gamma}_\nu = \nabla\times(\rho^{-1}\nabla\sigma) \,\,,
\ee
which is connected to the viscosity.

Considering initially that the system is simply composed of the electromagnetic field and a non-charged fluid, in this case the Lagrangian of the system is given by 

\be\label{01}
{\cal L} = {\cal L}_{Fluid} + {\cal L}_{Maxwell} = -{1\over 4}T_{\mu\nu}T^{\mu\nu}-{{1}\over{4}}F_{\mu\nu}F^{\mu\nu}\,\,.
\ee
where we have not considered the presence of source terms.

Let us now consider the case where the fluid is  charged, where the interaction between the fluid and the electromagnetic field  will be introduced inside this last Lagrangian by using a minimum coupling between the fluid's velocity field $(U^{(\epsilon)}_\mu)$ - here we have introduced the index $(\epsilon)$ valid for each species (positively or negatively charged), where the generalization to several species is straightforward - and the vector potential of the electromagnetic field $(A_\mu)$ is given by

\be\label{02}
U^{(\epsilon)}_\mu \longrightarrow U^{(\epsilon)}_\mu +g A_\mu \,\, .
\ee
Thus, we have that
\be\label{03}
T_{\mu\nu}  \longrightarrow T^{(\epsilon)}_{\mu\nu} =\partial_\mu U^{(\epsilon)}_{\nu} -\partial_\nu U^{(\epsilon)}_\mu \longrightarrow T^{(\epsilon)}_{\mu\nu} + g F_{\mu\nu}
\ee
which, substituting in (\ref{01}) gives us
\be\label{04}
{\cal L} = -{{1}\over{4}}T^{(\epsilon)}_{\mu\nu}T_{(\epsilon)}^{\mu\nu}-{{1}\over{4}}(1+g^2)F_{\mu\nu}F^{\mu\nu} -{{1}\over{2}}gT_{\mu\nu}F^{\mu\nu} \,\,,
\ee
where the coupling constant is $g=e_{\epsilon}/m_{\epsilon}$, $e_{\epsilon}$ is the charge and $m_{\epsilon}$ is the mass of the charge. The Euler-Lagrange equations of motion are
\be\label{05}
(1+g^2)\, \partial_\mu F^{\mu\nu} + g\partial_\mu T_{(\epsilon)}^{\mu\nu} =0 \,\,,
\ee

\ni and it is easy to see that (\ref{04}) is invariant under the gauge transformations,
$A_\mu \rightarrow A_\mu +\partial_\mu \Lambda$,
for the electromagnetic fields, and
$U^{(\epsilon)}_\mu \rightarrow U^{(\epsilon)}_\mu +\partial_\mu \Lambda$,
for the compressible fluid field. Concerning the potentials, $U^{(\epsilon)}_\alpha$ and $A_\alpha$, the above equation reads
\be\label{06}
(1+g^2)\[ {\Box} A_{\mu} - \partial_{\mu} \partial_{\nu} A^{\nu} \] + g\[ {\Box} U^{(\epsilon)}_{\mu} -\partial_{\mu}\partial_{\nu} U_{(\epsilon)}^{\nu} \]=0 \,\,.
\ee

\ni From now on, for simplicity, we will not use the species index, and much of what follows is true for each species.
The last term in (\ref{04}) is exactly the interaction between the charged fluid and the electromagnetic field applied. We can also observe that when the coupling constant is zero (the fluid is non-charged) we can obtain the Lagrangian (\ref{01}) again.

We can rewrite  the Lagrangian in Eq. (\ref{04}) as
\be\label{06.1}
{\cal L} = -{{1}\over{4}}T_{\mu\nu}T^{\mu\nu}-{{1}\over{4}}(1+g^2)F_{\mu\nu}F^{\mu\nu} -gU_{\mu}\partial_\nu F^{\mu\nu} \,\,,
\ee
where now we have a higher-derivative in the Maxwell sector represented by the last term in  (\ref{05}). Hence, there should be introduced another set of canonical pair $(\Sigma^{\mu}=\partial_0 A^{\mu}, \phi_\mu)$ to have a correct expanded phase space in order to proceed with the canonical analysis. Consequently, one can find the following Lagrangian 
\ba\label{06}
{\cal L} &=& {1\over2}( \dot{\vec U} -\nabla U_0)^2 +{1\over 4} a_{0}^2(\nabla \times \vec{U})^2 + {1\over2}(1+g^2) (\vec \Sigma -\nabla A_0 )^2  + {1\over4}(1+g^2) ( \nabla \times \vec A )^2 \nonumber\\ 
&-&g\vec U \cdot (\nabla \Sigma^{0} -\dot{\vec \Sigma}) -g U_0 (\nabla \cdot \vec \Sigma -\nabla^2 A_0 ) -g \vec U \cdot ( \nabla \times\nabla\times \vec{A})\,\,.
\ea
and the canonical Hamiltonian of the theory $H_c$ can be written as
\be\label{07}
H_c =\int d^3 x (p_{\mu}{\dot U}^{\mu} + \pi_{\mu}\dot{A}^{\mu} +\phi_{\mu}\dot{\Sigma}^{\mu} - {\cal L} ) \,\,,
\ee
where the momenta canonically conjugated to the fields $U^{\mu}$, $A^{\mu}$ and $\Sigma^{\mu} \equiv \dot{A}^{\mu}$, which can be considered as  independent variables, defined respectively by
\ba\label{08}
p_{\mu} &\equiv& {{\partial {\cal L}}\over{\partial(\dot{U}^\mu)}} \;,\\
\pi_{\mu} &\equiv& {{\partial {\cal L}}\over{\partial(\dot{A}^\mu)}} -\partial_0 \left[ {{\partial {\cal L}}\over{\partial(\ddot{A}^\mu)}}\right] -\partial_k\left[ {{\partial {\cal L}}\over{\partial(\partial_0\partial_k{A}^\mu)}} + {{\partial {\cal L}}\over{\partial(\partial_k{A}^\mu)}}\right] \;,\\
\phi_\mu &\equiv& {{\partial {\cal L}}\over{\partial(\ddot{A}^\mu)}}\;,
\ea
which result in the following expressions
\ba\label{09}
p_\mu &=& T_{\mu 0} \,\,, \nonumber \\
%\quad , \quad 
\phi_\mu &=& g\eta_{\mu k}U_k \,\,,\\
\pi_\mu&=&(1+g^2)F_{\mu 0} -g\eta_{\mu k}T_{0k} +g\eta_{\mu 0}\partial_k U_k \,\,.\nonumber
\ea

So, using the equations (\ref{08})-(\ref{09}) we can write the canonical Hamiltonian density ${\cal H}_c$ as
\ba\label{11}
{\cal H}_c &=& \pi_\mu \Sigma^{\mu} +{1\over2}{\vec{p}}^2 +\vec{p}\cdot\nabla U_0 -{1\over 2}a_0^2 (\nabla\times\vec{U} )^2 -{1\over2}(1+g^2)(\vec\Sigma -\nabla A_0 )^2\nonumber\\ &-&{1\over2}(1+g^2)(\nabla\times\vec{A})^2 +g \vec{U}\cdot\nabla\Sigma^{0} + gU_0(\nabla\cdot\vec{\Sigma}-\nabla^2 A_0 ) + g\vec{U}\cdot(\nabla\times\nabla\times\vec{A}) \,\,.
\ea

Therefore,  by working out a pure Dirac analysis of the Hamiltonian (\ref{11}) \cite{Dirac}, we notice that by the equations (\ref{09}) we have obtained the set of constraints 
\be\label{12}
\chi_1 \equiv \pi_0 -\nabla \cdot\phi \approx 0 \;,
\ee
\be\label{13}
\chi_2 \equiv \phi_0 \approx 0\;,
\ee
\be\label{14}
\chi_3 \equiv \nabla\cdot\vec\pi\approx 0\;,
\ee
\be\label{15}
\chi_4 \equiv p_0 \approx 0\;,
\ee
\be\label{16}
\chi_5 \equiv \nabla \cdot\vec{p} -g(\nabla \cdot\vec{\Sigma} -\nabla^2 A_0 )\approx 0 \;,
\ee
where $\chi_3$ was obtained by the time evolution of $\chi_1$, and $\chi_5$, by the time evolution of $\chi_4$. The set of constraints $\chi_{i},\;\; i=1,...,5$ are clearly first-class and no more new constraints are obtained.
With a usual notation, the symbol ``$\approx$'' means  {\it weak equality}. Following  Dirac's procedure, we have to choose five gauge conditions.

These conditions can be suggested by many reasons and the most important one can be the way it may simplify the theory. In Maxwell's theory, the condition usually employed to be the gauge fixing is the Coulomb gauge
\be\label{16.1}
\partial_\alpha A^\alpha = 0\,\,.
\ee
However, concerning the theory described by the Lagrangian in Eq. (\ref{01}), where a charged compressible fluid is immersed in an electromagnetic field, the condition (\ref{16.1}) is not sufficient to promote the mentioned gauge fixing.  In order to do that we need an extra condition. In this case, an appropriate choice could be the  ``Lorentz Gauge" for a compressible fluid \cite{albert1}, where
\be \label{16.2}
\partial_\alpha U^{\alpha} =0 \,\,,
\ee
which is directly related to the condition relative to the compressibility of the fluid \cite{albert1}. Thus, considering Eqs. (\ref{16.1}) and (\ref{16.2}) we can obtain the following set of gauge conditions
\be\label{17}
\Phi_1 \equiv A_0 \approx 0\;,
\ee
\be\label{18}
\Phi_2 \equiv \Sigma_0 \approx 0 \,\,\, (\dot{A}_0 \approx 0)\;,
\ee
\be\label{19}
\Phi_3 \equiv \nabla \cdot{\vec{A}} \approx 0\;,
\ee
\be\label{20}
\Phi_4 \equiv U_0 - \alpha \approx 0\;,
\ee
\be\label{21}
\Phi_5 \equiv \nabla \cdot \vec{U} \approx 0 \,\,,
\ee
where $\alpha$ in Eq. (\ref{17}) is a constant.
This set of constraints constitutes an appropriated noncovariant gauge condition which fixes the first-class constraints. 

%%%%%%%%%%%%%%%%%%%%%%%%%%%%%%%%%%%%%%%%%%%%%%%%%%%%%%%%%%%%%%%%%%%%%%%%%%%%%%%%%%%%%%%%%%%%%%%%%%%%%%%%%%%%%%%%%%%%%%%%%%%%%%%%%%%%%%%%%%%

\section{Path Integral Formalism}

Now, we are able to write down the generating functional, or transition amplitude, 
\be\label{22}
Z= \int Dp_\nu DU^{\nu}D\phi_\nu D\Sigma^{\nu} D\pi_\nu DA^\nu det\{\chi_a , \Phi_b \}\left[ \prod_{n=1}^{5} \delta[\chi_n]\delta[\Phi_n]\right]exp\left(i\int d^4 x \,{\cal L}_c\right) \,\,,
\ee
where the determinant between the first-class constraints (\ref{12})-(\ref{16}) and the gauge-fixing condition (\ref{17})-(\ref{21}) has the form
\be\label{23}
det\{\chi_a , \Phi_b \}=det\[\nabla^4 \]\;,
\ee
which does not contain any field variables and it can be put within a normalization constant.  Hence,

\be\label{24}
{\cal L}_c = p_\mu \partial_t U^\mu + \pi_\mu \partial_t A^\mu + \phi_\mu \partial_t \Sigma^\mu - {\cal H}_c \,\,.
\ee

Introducing Eqs. (\ref{11}), (\ref{12})-(\ref{16}), (\ref{17})-(\ref{21}), (\ref{23}) and (\ref{24}) into (\ref{22}), integrating over the momenta and field variables and, using the delta functional, we can obtain the following expression for the transition amplitude 
\be\label{25}
Z= \int Du^\mu DA^\mu \delta(\partial_k U^k)\delta(\partial_s A^s )exp\left( i\int d^4 x\, {\cal L}\right) \,\,,
\ee
where $\cal{L}$ is given by Eq. (\ref{05}).

Using a straight-forward generalization of the Faddeev-Popov ansatz, we can go from a noncovariant gauge fixing form to a covariant one such as
\be\label{26}
Z= \int DU^\mu DA^{\mu}\delta\left[{1\over\xi}\partial_s U^s -f\right]\delta\left[{1\over\Lambda}\partial_s A^s -f^{\prime}\right]exp\left(i\int d^4 x \, {\cal L}\right) \,\,,
\ee
where $\xi \neq 0$ and $\Lambda \neq 0$ are arbitrary real numbers and $f=f(x)$ and $f^{\prime}=f^{\prime}(x)$ are arbitrary real functions. 

Now, since the generating functional is independent of  $f(x)$ and $f^{\prime}(x)$ we can integrate in $f(x)$ with weight $exp\left(-{i\over 2}\int d^4 x f^2\right)$ and in $f^{\prime}(x)$ with the weight $exp\left(-{i\over 2}\int d^4 x {f^{\prime}}^2\right)$
\ba\label{27}
\hat{Z}&=&\int Df Df^{\prime}Zexp\left(-{i\over 2}\int d^4 x f^2\right)exp\left(-{i\over 2}\int d^4 x {f^{\prime}}^2\right)\nonumber\\
&=&\int DcD\bar{c}\int DbD\bar{b}\int DU^{\mu}DA^{\mu} exp\left( i\int d^4 x \,{\cal L}_{eff}\right),
\ea
where $c,\,\bar{c}$ are the ghost fields by the Maxwell sector and $b,\,\bar{b}$ are the ghost fields for the fluid sector. The effective Lagrangian density ${\cal L}_{eff}$ is defined by
\be\label{28}
{\cal L}_{eff} = -{{1}\over{4}}T_{\mu\nu}T^{\mu\nu}-{{1}\over{4}}(1+g^2)F_{\mu\nu}F^{\mu\nu} -gU_{\mu}\partial_\nu F^{\mu\nu} -{1\over{2\xi^2}}(\partial_\mu U^{\mu})^2 -{1\over{2\Lambda^2}}(\partial_\mu A^{\mu})^2 -{1\over\xi}c\Delta\bar{c} -{1\over\Lambda}b\Delta\bar{b}\,\,,
\ee
where $\Delta =\partial_\nu \partial_\nu$\,.

In the next section we will analyze the theory in thermodynamic\l equilibrium and we will work with the partition function, which is the most important function in thermodynamics. From it, all the thermodynamical properties can be obtained, namely, pressure, particle numbers, entropy and energy.

%%%%%%%%%%%%%%%%%%%%%%%%%%%%%%%%%%%%%%%%%%%%%%%%%%%%%%%%%%%%%%%%%%%%%%%%%%%%%%%%%%%%%%%%%%%%%%%%%%%%%%%%%%%%%%%%%%%%%%%%%%%%%%%%%%%%%%%%%%%%%%%%%%%%%

\section{The Partition Function of Theory}

To obtain the partition function from the transition amplitude we have to carry out a kind of ``Euclideanization" of the time components of the vector  fields, a compactification  of the Wick-rotated time coordinate, and to impose periodic boundary conditions $(P)$ in this coordinate for the fluid, the electromagnetic  and ghost fields. Doing so, we can find the partition function
\be\label{29}
Z[\beta] = \int DcD\bar{c}\int DbD\bar{b} DU_\mu DA_\mu exp\left\{-\int_\beta dx \,{\cal L}_E\right\} \,\,,
\ee
where $\beta = 1/T$, $T$ is the temperature, and 
\be\label{30}
\int_\beta dx \equiv \int_{0}^{\beta} d\tau \int d^3 x\,\,,
\ee
and 
\be\label{31}
{\cal L}_{E} = -{{1}\over{4}}T_{\mu\nu}T^{\mu\nu}-{{1}\over{4}}(1+g^2)F_{\mu\nu}F^{\mu\nu} -gU^{\mu}\partial^\nu F_{\mu\nu} -{1\over{2\xi^2}}(\partial_\mu U_{\mu})^2 -{1\over{2\Lambda^2}}(\partial_\mu A_{\mu})^2 -{1\over\xi}c\Delta\bar{c} -{1\over\Lambda}b\Delta\bar{b} \,\,,
\ee
where ${\cal L}_E$ is the so-called effective Euclidean Lagrangian density.

So,  the partition function takes the form
\ba\label{32}
Z[\beta] &=& \int_P DcD\bar{c} \,\,exp\left\{\int_\beta dx\left[{1\over\xi}c\Delta\bar{c}\right]\right\}\int_P DbD\bar{b} \,\,exp\left\{\int_\beta dx\left[{1\over\Lambda}b\Delta\bar{b}\right]\right\}\nonumber\\
&\times&\int_P DU_\mu \,\,exp\left\{-{1\over 2}\int_\beta dx \,\, U^\mu {\cal O}_{\mu\nu}^{(f)} U^\nu \right\}\int_P DA_\mu \,\,exp\left\{-{(1+g^2)\over 2}\int_\beta dx A^\mu {\cal O}_{\mu\nu}^{(M)} A^\nu\right\} \nonumber\\
&\times& exp\left\{ \int_\beta dx \,\,gU^\mu\partial^\nu F_{\mu\nu}\right\},
\ea
where the operators ${\cal O}_{\mu\nu}^{(f)}$ (for the fluid sector) and ${\cal O}_{\mu\nu}^{(M)}$ (for the Maxwell sector) are defined by
\be\label{33}
{\cal O}_{\mu\nu}^{(f)} = \delta_{\mu\nu}\Delta -{{\xi^2 -1}\over\xi^2}\partial_\mu\partial_\nu
\ee
and
\be\label{34}
{\cal O}_{\mu\nu}^{(M)} = \delta_{\mu\nu}\Delta -{{\Lambda^2 -1}\over\Lambda^2}\partial_{\mu}\partial_\nu\,\,.
\ee

Note that the partition function in Eq. (\ref{32}) have a cross term (the last one in \eqref{32}) for the field of the fluid $(U_\mu)$ and the Maxwell field $(A_\mu)$, which does not allow a direct calculation.   It is independent of the functional integral concerning these fields, such as the fields for the ghosts that do not have the interaction for the fields $A_\mu$ and $U_\mu$, where
\be\label{35}
\int_P DcD\bar{c} \,\,exp\left\{\int_\beta dx\left[{1\over\xi}c\Delta\bar{c}\right]\right\} \equiv {\rm det}\left[{1\over\xi}\Delta\right]\,\,,
\ee
\be\label{36}
\int_P DbD\bar{b} \,\,exp\left\{\int_\beta dx\left[{1\over\Lambda}b\Delta\bar{b}\right]\right\} \equiv {\rm det}\left[{1\over\Lambda}\Delta\right]\,\,.
\ee

So, we will calculate the functional integral for the Maxwell field considering the cross term as an external source term for the electromagnetic field by doing the following transformation in the field $A_\mu$
\be\label{37}
A_\mu \longrightarrow A_\mu (1+g^2)^{-1/2} \,\,.
\ee
Thus we have
\ba\label{38}
&&\int_P DA_\mu \,\,exp\left\{-{(1+g^2)\over 2}\int_\beta dx A^\mu {\cal O}_{\mu\nu}^{(M)} A^\nu + \int_\beta dx\,\,A^\mu g\partial^\nu T_{\mu\nu}\right\} \nonumber\\ 
&=&\int_P DA_\mu \,\,exp\left\{-{1\over 2}\int_\beta dx A^\mu {\cal O}_{\mu\nu}^{(M)} A^\nu + \int_\beta dx\,\,A^\mu {g\over{(1+g^2)^{1/2}}}\partial^\nu T_{\mu\nu} \right\} =\nonumber\\
&=&\int_P DA_\mu \,\,exp\left\{-{1\over 2}\int_\beta dx A^\mu {\cal O}_{\mu\nu}^{(M)} A^\nu + \int_\beta dx\,\,A^\mu J_\mu^{(f)} \right\},
\ea
where $J_\mu^{(f)}$ (the above index $(f)$ means source term due to the coupling with the fluid) is defined as being
\be\label{39}
J_\mu^{(f)} =  {g\over{(1+g^2)^{1/2}}}\partial^\nu T_{\mu\nu}\,\,.
\ee
Hence, after integration in the gauge field $A_\mu$, we have that
\ba\label{40}
\int_P DA_\mu \,\,exp\,\bigglb\{-{1\over 2}\int_\beta dx A^\mu {\cal O}_{\mu\nu}^{(M)} A^\nu &+& \int_\beta dx\,\,A^\mu J_\mu^{(f)} \biggrb\}  \\
&=& \left[ {\rm Det}\left({\cal O}_{\mu\nu}^M\right) \right]^{-1/2}{1\over 2}\int_\beta dx\,\, J^\mu_{(f)}\left( {\cal O}_{\mu\nu}^{(M)}\right)^{-1}J^\nu_{(f)}\,\,, \nonumber
\ea
where ``Det'' in (\ref{40}) means the determinant in both Euclidean space-time and the Hilbert space. 

So, introducing (\ref{35}), (\ref{36}) , (\ref{39}) and (\ref{40}) into (\ref{32}) the  partition function is given by
\ba\label{41}
Z[\beta] &=& {\rm det}\left[{1\over\xi}\Delta\right]\,{\rm det}\left[{1\over\Lambda}\Delta\right]\,\left[ {\rm Det}\left({\cal O}_{\mu\nu}^M\right) \right]^{-1/2}\nonumber\\
&\times& \int_P\, DU_\mu \, exp\left\{ \int_\beta dx \left[-{1\over2}U^\mu{\cal O}_{\mu\nu}^{(f)} U^\nu +{1\over 2} {g^2\over{1+g^2}}U^\mu \left(\delta_{\mu\nu}\Delta -\partial_\mu \partial_\nu\right)U^\nu\right]\right\}
\ea
or, using the operator defined in (\ref{33}), we have after some steps that
\ba\label{42}
Z[\beta] &=& {\rm det}\left[{1\over\xi}\Delta\right]\,{\rm det}\left[{1\over\Lambda}\Delta\right]\,\left[ {\rm Det}\left({\cal O}_{\mu\nu}^M\right) \right]^{-1/2}\nonumber\\
&\times& \int_P\, DU_\mu \, exp\left\{ \int_\beta dx \left[-{1\over2}U^\mu\left[\left( 1 - {g^2\over{1+g^2}}\right)\delta_{\mu\nu}\Delta -\left({{\xi^2 -1}\over{\xi^2}}-{g^2\over{1+g^2}}\right)\partial_\mu\partial_\nu \right]U^\nu\right]\right\}\nonumber\\
&=&{\rm det}\left[{1\over\xi}\Delta\right]\,{\rm det}\left[{1\over\Lambda}\Delta\right]\,\left[ {\rm Det}\left({\cal O}_{\mu\nu}^M\right) \right]^{-1/2}\int_P\, DU_\mu \, exp\left\{ -{1\over2}\int_\beta dx \,U^\mu {\cal O}_{\mu\nu}U^\nu\right\},
\ea
where the operator ${\cal O}$ is defined by
\be\label{43}
{\cal O}_{\mu\nu} \equiv \left(1 - {g^2\over{1+g^2}}\right)\delta_{\mu\nu}\Delta -\left({{\xi^2 -1}\over{\xi^2}}-{g^2\over{1+g^2}}\right)\partial_\mu\partial_\nu.
\ee
Thus, we obtain that
\be\label{44}
Z[\beta] =  {\rm det}\left[{1\over\Lambda}\Delta\right]\,\left[ {\rm Det}\left({\cal O}_{\mu\nu}^M\right) \right]^{-1/2}\,{\rm det}\left[{1\over\xi}\Delta\right]\,\left[ {\rm Det}\left({\cal O}_{\mu\nu}\right) \right]^{-1/2}\,\,.
\ee
Since the temperature does not depend on $\xi$ or $\Lambda$, it can be included into the normalization constant.   Hence,  we can write the partition function after evaluating the determinant  in the Euclidean space-time  as
\be\label{45}
Z[\beta]= \left[ {\rm det} \,(\Delta) \right]^{-1} \,(1+g^2)^{-6}\,\left[ {\rm det}\, (\Delta) \right]^{-1}.
\ee                                                                                                            

We note that the partition function is a product of determinants of the form $[{\rm det}(\Delta + m_j^2)]^{(-n_j)/2}$, with $j=1$ and $2$. Each one of these terms describes a gas of free particles with mass $m_j$ and $n_j$ as being the degrees of freedom (DOF). We identify the first of these determinants as a partition function for massless particles with two DOF's, i.e., the Maxwell photons. On the other hand, the second determinant is the partition function for the fluid with two DOF's. In \cite{albert1}, in appendix A, we have made a brief discussion about the fluid DOF's.

In order to evaluate the determinants, we note that the equation
\be\label{46}
{\rm det}(\Delta) = \prod_{n, \vec p} \beta^2(\omega_n^2 +{\vec p}^{\,2})
\ee
and, using this identity, the logarithm of the partition function can be written as
\be\label{47}
\ln[Z(\beta)]= -\sum_{n, \vec p} \ln [\beta^2(\omega_n^2 +{\vec p}^2)] -\sum_{n, \vec p} \ln [(1+g^2)^{-6}\beta^2(\omega_n^2 +{\vec p}^2)] \,\,.
\ee
%
%or
%\be\label{48}
%\ln[Z(\beta)]= -\underbrace{\sum_{n, \vec p} \ln [\beta^2(\omega_n^2 +{\vec p}^2)]}_{Maxwell}- \underbrace{\sum_{n, \vec p} \ln [\beta^2(\omega_n^2 +{\vec p}^2)]}_{Fluid} +\underbrace{6\ln (1+g^2)}_{correction\,\, due\,\, to\,\, interaction}
%\ee

Now, evaluating the sum in $n$, and passing to the continuous in momentum space, we have that
\be\label{48}
\ln[Z(\beta,V] = -2V\int {{d^3\,p}\over{(2\pi)^3}}\ln (1-e^{-\beta p})\Big|_{\rm{Maxwell}} -2V\int {{d^3\,p}\over{(2\pi)^3}}\ln (1-e^{-\beta^{'} p})\Big|_{\rm{Fluid}} \,\,,
\ee
where $\beta^{'} =\beta(1+g^2)^{-3}$, and we find
\be\label{49}
\ln[Z(\beta,V] ={{\pi^2}\over{45}}{{V}\over{\beta^3}} + {{\pi^2}\over{45}}{{V}\over{\beta^{'3}}} = {{\pi^2}\over{45}}{{V}\over{\beta^3}} + {{\pi^2}\over{45}}{{V}\over{\beta^3}}(1+g^2)^9 \,\,.
\ee

The first term in Eq. (\ref{49}) is the usual Planck result associated with the free Maxwell and it gives the usual Stefan-Boltzmann law. The second term corresponds to the fluid Stefan-Boltzmann type law \cite{Lima}. This is a very interesting result, which allows us to derive some properties of our system. Now, starting from (\ref{49}), which can be rewritten as
\be\label{50}
\ln[Z(\beta,V] ={{\pi^2}\over{45}}{{V}\over{\beta^3}}[1 + (1+g^2)^9] \,\,,
\ee
we can obtain the energy density defined by
\ba\label{51}
\rho &=& {{k_B T^2}\over{V}} {{\partial}\over{\partial T}}\ln[Z(\beta,V]\nonumber\\
&=& k_B {{\pi^2}\over{15}}T^4 [1+ (1+g^2)^9] \,\,,
\ea
where $k_B$ is the Boltzmann constant, and the pressure defined by 
\ba\label{52}
p&=&k_B T {{\partial}\over{\partial V}}\ln[Z(\beta,V]\nonumber\\
&=&{{1}\over{3}}k_B {{\pi^2}\over{15}}T^4 [1+ (1+g^2)^9].
\ea

Notice that we can obtain a relationship between these quantities given by
\be\label{53}
p={{1}\over{3}}\rho(g) \,\,,
\ee
which is a equation of state, that depends on the coupling constant $g$.

It is also interesting to note that is a particular case of the ``gamma-law'' equation of state 
\be\label{54}
p=(\gamma -1)\rho\,\,,
\ee
where the index $\gamma =4/3$ that corresponds to a radiation fluid \cite{Lima}. Therefore, from (\ref{51}) we have a generalization of the Stefan-Boltzmann law, given by
\be\label{55}
\rho(g) =\eta(g) T^4\,\,,
\ee
where $\eta$ depends on the coupling constant $g$.  We will comment this result in the next section.

\section{Conclusions}

The current fluid dynamics literature has several motivations that keep the interest in this subject at high levels during this last decades.   The interaction with the electromagnetic fields is one of these motivations.

In this work we have provided a constraint analysis of the Lagrangian system formed by this kind of interaction.  The final Lagrangian has higher derivatives five first-class constraints.  The Lorentz gauge fixing for compressible fluids was used and the set of gauge transformations was described.

We have used statistical elements such as the partition function, and the Faddeev-Popov ansatz, to obtain an equation of state similar to the dark energy model for the accelerated Universe one. The final result is the fourth power temperature dependence, which shows a direct analogy to the Stefan-Boltzmann law, where the coefficient term is a function of the coupling constant.

\section{Acknowledgments}

\ni The authors thank CNPq (Conselho Nacional de Desenvolvimento Cient\' ifico e Tecnol\'ogico),
Brazilian scientific support federal agency, for partial financial support, Grants
numbers 302155/2015-5, 302156/2015-1 and 442369/2014-0 and E.M.C.A. thanks the hospitality of Theoretical Physics Department at Federal University of Rio de Janeiro (UFRJ), where part of this work was carried out.

\end{document}